# BatMan: Mitigating Batch Effects via Stratification for Survival Outcome Prediction

**AI NI**

Division of Biostatistics, College of Public Health, Ohio State University, Columbus, OH 43210

**MENGLING LIU**

Department of Population Health, New York University, New York, NY 10016

**LI-XUAN QIN**

Department of Epidemiology and Biostatistics, Memorial Sloan Kettering Cancer Center, New York, NY 10017, Email: qinl@mskcc.org

# ABSTRACT

Reproducible translation of transcriptomics data has been hampered by the ubiquitous presence of batch effects. Statistical methods for managing batch effects were initially developed in the setting of sample group comparison and later borrowed for other settings such as survival outcome prediction. The most notable such method is ComBat, which adjusts for batches by including it as a covariate alongside sample groups in a linear regression. In survival prediction, however, ComBat is used without definable groups for survival outcome and is done sequentially with survival regression for a potentially confounded outcome. To address these issues, we propose a new method, called BatMan ("BATch MitigAtion via stratificatioN"). It adjusts batches as strata in survival regression and utilize variable selection methods such as LASSO to handle high dimensionality. We assess the performance of BatMan in comparison with ComBat, each used either alone or in conjunction with data normalization, in a re-sampling-based simulation study under various levels of predictive signal strength and patterns of batch-outcome association. Our simulations show that (1) BatMan outperforms ComBat in nearly all scenarios when there are batch effects in the data, and (2) their performance can be worsened by the addition of data normalization. We further evaluate them using microRNA data for ovarian cancer from the Cancer Genome Atlas, and find that BatMan outforms ComBat while the addition of data normalization worsens the prediction. Our study thus shows the advantage of BatMan and raises caution about the naïve use of data normalization in the context of developing survival prediction models. The BatMan method and the simulation tool for performance assessment are implemented in R and publicly available at https://github.com/LXQin/PRECISION.survival.

# 1. INTRODUCTION

Recent decades have seen accumulating transcriptomics data produced by high-throughput technologies such as microarrays and sequencing (Consortium *et al.*, 2006, Consortium, 2014, Mestdagh *et al.*, 2014). Their translation into useful biomedical knowledge, however, has been complicated by the ubiquitous presence of batch effects, namely systematic data variations between data collection batches that undergo varied experimental handling conditions such as regents, equipment, and personnel (Irizarry *et al.*, 2003, Kratz and Carninci, 2014, Leek *et al.*, 2010). Failure to properly manage batch effects has led to subpar and irreproducible interpretations of the data (Brown, Kaiser and Allison, 2018, Loven *et al.*, 2012, Nekrutenko and Taylor, 2012). Statistical methods for managing batch effects have been developed in the setting of differential expression analysis for group comparison (Gagnon-Bartsch and Speed, 2012, Johnson, Li and Rabinovic, 2007, Lazar *et al.*, 2013, Leek and Storey, 2007). The predominantly popular method is ComBat, which uses an Empirical Bayes approach to pool feature-specific estimates of batch effects across features (Johnson, Li and Rabinovic, 2007). Despite its popularity, ComBat has recently been shown to be unreliable when data batches and sample groups are confounded (Li *et al.*, 2021, Nygaard, Rodland and Hovig, 2016). In this case, it may not only induce bias in the estimated group mean differences for individual features but also inflate correlations among molecular features (due to the deduction of estimated batch effects that are shared by features), leading to negative impact on downstream analysis.

Since its initial development, ComBat has also been used in the setting of survival outcome prediction – an important goal in biomedical studies for disease management and trial design (Lee, 2005, Li *et al.*, 2010, van't Veer and Bernards, 2008). In this setting, transcriptomics data are first adjusted for batches using ComBat without sample groups and then used as covariates in survival regression (Han, Sayyid and Altman, 2021, McKinney *et al.*, 2021, Wang *et al.*, 2017). As shown in this article, such sequential operation of batch adjustment and survival analysis is inadequate, especially when batches and survival outcome are confounded. Therefore, we propose a new method for managing batch effects in survival prediction. Our method enables concurrent batch management and survival prediction by adjusting batches as strata in Cox proportional hazards regression (Kalbfleisch and Prentice, 2002). In stratified Cox regression models, the baseline hazard is stratum-specific and a common hazard ratio for each covariate is aggregated over strata (Bradburn *et al.*, 2003, Cochran, 1968). It has been previously used in low-dimensional data to handle covariates that deviate from the proportional hazard assumption and to estimate the interaction between a covariate and the stratification variable (Altman, 1990). For managing batch effects in survival analysis of high-dimensional transcriptomics data, we propose to combine batch stratification and variable selection to simultaneously mitigate batch effects and build survival predictors. We call this new method BatMan ("BATch MitigAtion via stratificatioN").

To investigate the performance of BatMan in comparison with ComBat for survival prediction, we conduct a simulation study using data simulated from a pair of microRNA (miRNA) microarray datasets (Bartel, 2004, Qin *et al.*, 2014, Qin *et al.*, 2018). Two datasets were previously collected for the same set of tumor samples, where arrays in one dataset were processed with uniform handling in a single batch and arrays in the other dataset were generated with disparate handling in multiple batches (Qin *et al.*, 2014, Qin *et al.*, 2018). Leveraging the paired datasets, we developed a re-sampling-based approach for simulating array data and a permutation-based procedure for generating outcome data under prespecified levels of association (Ni and Qin, 2021, Qin, Huang and Begg, 2016). Here we use this approach to generate data under various scenarios for assessing the performance of BatMan and ComBat, each either applied alone or together with data normalization (Irizarry *et al.*, 2003). Our simulations show that (1) BatMan outperforms ComBat in nearly all scenarios when there are batch effects in the training data or test data, and (2) their performance can be worsened by the addition of data normalization.

We further evaluate the performance of BatMan and ComBat using the miRNA array data for ovarian cancer from the Cancer Genome Atlas (TCGA) (Cancer Genome Atlas Research, 2011). Ovarian cancer is the fifth most common cause of cancer deaths in women in the United States (ACS, 2022). The majority of ovarian carcinomas are of high-grade serous histology and associated with poor prognosis; nearly one-third of these patients will not respond to initial treatment with surgery and chemotherapy and die within one year of diagnosis (Barlin *et al.*, 2012). Molecular predictors are thus needed to improve patient stratification and treatment options in this challenging disease (Zhang *et al.*, 2020). This analysis shows that (1) BatMan leads to more accurate outcome prediction than ComBat, and (2) the prediction accuracy is worsened by the addition of data normalization.

## 2. METHODS

### 2.1 Batch Mitigation via Stratification (BatMan)

We propose to use the stratified Cox regression model in combination with variable selection methods to build a model for survival probability prediction while mitigating the negative impact of batch effects. Suppose there are $B$ batches in the data for $n$ samples. Let $T_{bi}$ and $C_{bi}$ be the time to the outcome of interest and the censoring time, respectively, for sample $i$ in batch $b$ ($i = 1, \ldots, n_b, b = 1, \ldots, B$). Let $\Delta_{bi} = I(T_{bi} \leq C_{bi})$ be the censoring indicator, where $I(\cdot)$ denotes the indicator function. Suppose $G$ molecular features are measured for each sample. Further denote the $G$-dimensional true expression levels of measured molecular features for sample $i$ in batch $b$ by $X_{bi}$ and the vector of batch effects by $W_b = (W_{1b}, \ldots, W_{Gb})$, where $W_{gb}$ ($g = 1, \ldots, G$) is the feature-specific batch effect for feature $g$ in batch $b$. Thus, we allow batch effects to vary across features and shared across samples in the same batch. The observed expression levels that are contaminated by batch effects can be written as $Z_{bi} = X_{bi} + W_b$. Let $\beta_0$ be the true $G$-dimensional regression coefficients of $X$. Let $h_{b0}(t)$ be the batch-specific baseline hazard function of the outcome event. Under the proposed batch-stratified Cox regression model,

$$h(t \mid Z_{bi}) = h_{b0}(t)\exp(Z'_{bi}\beta).$$

Suppose there are $m_b$ unique event times in batch $b$, denoted by $t_{bi}$ ($i = 1, \ldots, m_b$). Denote the batch-specific risk set at time $t_{bi}$ by $R(t_{bi})$. The partial likelihood function for the batch-stratified Cox model is

$$L_s(\beta) = \prod_{b=1}^{B} \prod_{i=1}^{m_b} \frac{\exp(Z'_{bi}\beta)}{\sum_{j \in R(t_{bi})} \exp(Z'_{bj}\beta)}$$

$$= \prod_{b=1}^{B} \prod_{i=1}^{m_b} \frac{\exp((X'_{bi} + W'_b)\beta)}{\sum_{j \in R(t_{bi})} \exp((X'_{bj} + W'_b)\beta)}$$

$$= \prod_{b=1}^{B} \prod_{i=1}^{m_b} \frac{\exp(X'_{bi}\beta)}{\sum_{j \in R(t_{bi})} \exp(X'_{bj}\beta)}$$

Thus, it is clear that batch effects cancel out due to the structure of the stratified partial likelihood function and the, by definition, batch-specific bath effects. In this article, we consider a batch to be either an array slide (containing multiple arrays) or an experimental batch (containing multiple array slides).

Since $G$ is typically much larger than the sample size $n$ in a transcriptomic study, a variable selection method is needed to build a parsimonious model of features that are strongly associated with the survival outcome. We implement the stratified Cox regression with two commonly used methods for variable selection: (1) univariate variable selection based on p-values and (2) regularized regression with a penalty term (Hastie *et al.*, 2009).

- In the univariate method, we assess the significance of survival outcome association for each feature using a univariate stratified Cox regression model and select features with a p value less than or equal to a given cutoff. The selected features are then included in a multivariate stratified Cox model. The p value cutoff is selected via a grid search (for example, from 0 to 0.01 by 0.0005) to minimize the Akaike information criterion (AIC) in the multivariate model (Akaike, 1973).

- In the regularized method, the regression coefficient $\beta$ is estimated based on the penalized logarithm of the stratified partial likelihood function in the following form:

$$\log(L_s(\beta)) - n\lambda \sum_{g=1}^{G} J(\beta_g),$$

where $\lambda$ is a tuning parameter that controls the magnitude of the penalty and $\beta_g$ is the $g$-th component of $\beta$. In this article, we focus on the use of the Lasso penalty where $J(\beta_g) = |\beta_g|$ and the adaptive Lasso penalty where $J(\beta_g) = |\beta_g|/|\tilde{\beta}_g|$, with $\tilde{\beta}_g$ being a consistent estimator of the true value $\beta_{0g}$ (Tibshirani, 1997). The adaptive Lasso penalty is a modified version of the Lasso penalty that renders it the oracle property (Zou, 2006). They both can provide sparse estimates of $\beta$ with some components being exactly zero, thereby achieving variable selection. Cross-validation is used to select the tuning parameter $\lambda$ via grid search. The coordinate descent algorithm is modified by incorporating stratification and then used for estimating the regularized regression coefficients (Simon *et al.*, 2011). To speed up the computation, we precede fitting the regularized regression model with a variable screening step, which uses the univariate stratified Cox model to select the top $[n_0/4]$ features with the largest univariate partial likelihood, where $n_0$ is the number of events in the data and [.] denotes the nearest integer. This two-step variable selection strategy has been extensively studied in high-dimensional data analysis literature and showed satisfactory performance (Fan, Samworth and Wu, 2009, Fan and Lv, 2008, Tamba, Ni and Zhang, 2017).

**2.2 Empirical Bayes Method for Adjusting Batch Effects (ComBat)**

This section provides a brief overview of the ComBat method (Johnson, Li and Rabinovic, 2007). ComBat adopts a location and scale model. Following similar notations as in BatMan, the model underlying ComBat can be written as:

$$Z_{big} = \alpha_{0g} + V\alpha_{1g} + \gamma_{bg} + \delta_{bg}\varepsilon_{big},$$

Where $Z_{big}$ is the observed expression level of feature $g$ in sample $i$ from batch $b$, $\alpha_{0g}$ is the baseline expression level of feature $g$, and $\alpha_{1g}$ is the regression coefficient vector corresponding to the design matrix $V$ that contains covariates thought to be associated with feature expression levels. Note that we do not consider any such covariates in this study. The batch-related parameters $\gamma_{bg}$ and $\delta_{bg}$ represent the additive and multiplicative batch effects of batch $b$ for feature $g$, respectively. The error term $\varepsilon_{big}$ is assumed to follow a normal distribution with mean zero and variance $\sigma_g^2$.

The first step of ComBat is to calculate the standardized data $S_{big}$ that only contains information on batch effects:

$$S_{big} = \frac{Z_{big} - \hat{\alpha}_{0g} - V\hat{\alpha}_{1g}}{\hat{\sigma}_g},$$

where $\hat{\alpha}_{0g}$, $\hat{\alpha}_{1g}$ and $\hat{\sigma}_g$ are estimated from the feature-wise univariate ordinary linear regression with intercept, $V$, and batch indicator. Based on a location-and-scale model, $S_{big}$ follows the $N(\gamma_{bg}^*, \delta_{bg}^2)$ distribution, where $\gamma_{bg}^* = \gamma_{bg}/\hat{\sigma}_g$. Next, $\gamma_{bg}^*$ and $\delta_{bg}^2$ are assumed to follow prior distributions of $N(\gamma_b, \tau_b^2)$ and Inverse Gamma $(\lambda_b, \theta_b)$, respectively. Hyperparameters $\gamma_b, \tau_b^2, \lambda_b, \theta_b$ are estimated empirically from the standardized data using the method of moments. When there is evidence that the normal and inverse gamma prior distributions assumed for $\gamma_{bg}^*$ and $\delta_{bg}^2$, respectively, do not fit the data well, a non-parametric prior method can be used instead. Finally, the Empirical Bayes estimates of parameters, $\hat{\gamma}_{bg}^*$ and $\hat{\delta}_{bg}^2$, are used to calculate the batch-adjusted data $Z_{big}^*$:

$$Z_{big}^* = \frac{\hat{\sigma}_{0g}}{\hat{\delta}_{bg}}\left(S_{big} - \hat{\gamma}_{bg}^*\right) + \hat{\alpha}_{0g} + V\hat{\alpha}_{1g}.$$

The batch-adjusted data are then used for subsequent analyses such as survival prediction.

## 2.3 Simulation Study

*Collection of the empirical data*      Details on data collection can be found in Qin et al. ( Qin *et al.*, 2014, Qin, Huang and Begg, 2016, Qin *et al.*, 2018). Briefly, 96 high-grade serous ovarian cancer samples and 96 endometroid endometrial cancer samples were collected at Memorial Sloan Kettering Cancer Center between 2000 and 2012. Their miRNA expression levels were measured twice using Agilent microarrays (Release 16.0, Agilent Technologies, Santa Clara, CA), one with uniform handling that minimized batch effects and the other with typical practices where the arrays were processed in multiple batches resulting in batch effects. We call the first design the 'uniformly-handled' design and the second the 'non-uniformly-handled' design. The raw array data were preprocessed with log2 transformation and median summarization across replicate probes for each feature (Qin, Huang and Zhou, 2014).

*Estimation of biological effects*      We use the data collected with the uniformly handled design from the 96 ovarian-cancer samples as a best approximate for the biological effects of these samples. We call them 'virtual samples'.

*Estimation of handling effects* Assuming that handling effects are additive, we use the differences between the uniformly handled and the non-uniformly handled datasets to estimate the handling effects of the arrays in the latter dataset. We call them 'virtual arrays' and split them (by whole slides, each of which contains eight arrays) into two 96-array sets, one for prediction model training and another for validation. The additivity assumption has been deemed reasonable for microarray data and has been adopted in published methods on microarray data normalization and analysis (Kerr, Martin and Churchill, 2000, Qin and Satagopan, 2009).

*Elicitation of regression coefficients* We use the virtual samples to assess each feature's association with progression free survival (PFS), an important survival outcome variable in ovarian cancer (Barlin *et al.*, 2012). PFS is defined as the time from primary surgery to disease progression, death, or loss of follow-up, whichever occurs first. The censoring rate was 23% (22/96) in the data. Univariate Cox regression analysis identifies six features with p values less than 0.005 whose log hazard ratio ranged from 0.26 to 0.78 (Qin and Levine, 2016). Three sets of 'true regression coefficients' from the estimated regression coefficients are generated with various signal levels: (1) Moderate signal, which uses the quadruple of the univariately estimated log hazard ratios for the six most sigificant features and zero for the other features; (2) Weak signal, which uses 0.35 for 30 randomly selected features (irrespective of the six features used in moderate signal scenario and fixed selection for all simulation replications), and zero for the other features; (3) Null signal, which uses zero for all features. We set the L1 norm of the moderate-signal coefficient vector to be equal to that of the weak-signal coefficient vector, so that the two vectors have the same "total" effect that is distributed differently across features.

*Simulation of survival outcome* We use a previously developed permutation-based method to simulate the new PFS outcome based on the true regression coefficients and biological effects of features. The simulated PFS maintains the same marginal distribution as the observed PFS. Briefly, the observed PFS times are sorted in ascending order and then sequentially and probabilistically paired with the virtual samples based on multinomial distributions whose parameters are the partial likelihood of each unpaired virtual sample assuming it fails at the given PFS time. Detailed description of this algorithm can be found in Ni and Qin (2021).

*Simulation of miRNA data*   To avoid overfitting, we simulate both training data and test data to build prediction models and evaluate their prediction performance, respectively. Training and test data are both simulated with virtual re-hybridization to preserve the complex across-feature correlation structure of biological effects and handling effects (Qin, Huang and Begg, 2016, Qin *et al.*, 2014). Namely, the 96 virtual samples (along with their PFS times) are reassigned to the 96 virtual arrays allocated for training and test, respectively. Handling effects for each virtual array are then added to biological effects of the assigned virtual sample. We consider two scenarios for the reassignment: (1) randomized, and (2) sorted by simulated PFS time before reassignment so that PFS times are associated with handling effects.

*Batch effects adjustment*   We adjust handling effects that can be attributed to the data collection batches in the simulated data using BatMan or ComBat, either alone or following data normalization (median normalization, quantile normalization, and variance stabilizing normalization). ComBat is applied using the *ComBat* function in the R package *sva*; quantile normalization is applied using the *normalize.quantiles* function in the R package *preprocessCore*; variance stabilizing normalization uses the *vsn2* function in the R package *vsn*. Similar methods are applied to the training data and test data, except that the frozen version of a corresponding normalization method is used for the test data (McCall, Bolstad and Irizarry, 2010). That is, frozen quantile normalization is used for test data where quantiles derived from the training data are used to normalize the test data; VSN parameters that are estimated in the training data are used to normalize the test data.

*Prediction model training*   To reduce computational burden in simulations and alleviate collinearity in model fitting, we pre-filter the features using two criteria: (1) high abundance (mean expression on the log2 scale among the 96 samples >=8), and (2) no strong inter-features correlation (Pearson correlation coefficient <0.9). These criteria are applied to each simulated dataset, resulting in varied sets of features that pass the filtering across simulation runs. Typically, 100 to 200 features are kept for the model-training step. The same two methods for variable selection (univariate variable selection and regularized regression) introduced for BatMan are also used for the regression model fitting following ComBat. As a reference, we add a third method, which contains the true predictive features, referred to as *the oracle method*. Although the oracle model is not obtainable in practice, these results are nevertheless helpful for assessing the impact of batch effects and the effectiveness of batch effects adjustment methods on prediction accuracy.

*Prediction model evaluation*  Harrell's C-index is used to evaluate prediction accuracy (Harrell, Lee and Mark, 1996). It is a function of the test data and regression coefficients of a trained model. For stratified Cox models, C-index is calculated as the weighted average of stratum-specific C-indices with the weight being the number of comparable pairs in each stratum. The mean, standard deviation, $2.5^{th}$, and $97.5^{th}$ percentile of the C-index among the 400 runs for each simulation scenario are calculated.

*Simulation scenarios*       **Table 1** summarizes the eight scenarios investigated in our simulations. The scenarios are arranged in such a way that batch effects are increasingly prevalent and involved (that is, associated with the survival outcome).

To summarize, our simulations assess the performance of 19 approaches for batch-effects management in combination with 4 methods for prediction model development under 8 scenarios for data simulation.

**2.4 Application to TCGA Ovarian Cancer Study**

The proposed BatMan method, in comparison with the ComBat method, is applied to miRNA microarray data from the TCGA ovarian cancer study (Cancer Genome Atlas Research, 2011). The miRNA data along with the PFS data for 466 patients with stage II to IV primary, high-grade, untreated ovarian cancer were previously downloaded from the TCGA data portal (Hutter and Zenklusen, 2018). Among the 466 patients, 342 (73.3%) had disease progression or died during the study follow up. The batch variable, defined as the first shipment from the Biospecimen Core Resource (BCR), were downloaded at the TCGA Batch Viewer website (Akbani *et al.*, 2022). The same strategies for managing batch effects and building prediction models in our simulation study are applied to build a prediction model for PFS using the TCGA miRNA data. The 466 samples are split, by whole batches, into four folds, with three folds used for model training and the remaining fold for testing as measured by the C-index; this procedure is repeated 50 times and the average C-index is reported.

## 3. RESULTS

**3.1 Simulation with Moderate Signals**

We present the results of the oracle, LASSO-penalized, and univariate-selection methods in the main text. In addition, we include in Supplementary Material the results of the adaptive-LASSO-penalized method, which is very similar to that of the LASSO-penalized method (**S-Figure 1**).

*Oracle method*

**Figure** 1 shows the simulation results for the oracle method in terms of the prediction accuracy measured by the test-data C-index.

When batch effects are absent (BE00Cor00), BatMan (alone) leads to the best prediction accuracy when either batch- or slide-stratification is used (C-index 0.94 and 0.95, respectively); ComBat by batches is nearly as good (C-index 0.94), while ComBat by slides is slightly worse (C-index 0.90); the three normalization methods, when used alone, have varied accuracy, with median normalization being close to BatMan and ComBat (C-index 0.92) and VSN and quantile normalization significantly worse (C-index 0.80 and 0.74, respectively); the combinations of normalization with BatMan or ComBat show similar accuracy to that of normalization alone (C-index 0.73~0.92).

When batch effects are present but not associated with survival time in the test data (BE10Cor00, BE10Cor10, BE11Cor00, and BE11Cor10), BatMan by slides offers notably better accuracy (C-index 0.79~0.89) than any other method under investigation. Its performance worsens slightly when used with median normalization (C-index 0.78~0.88). BatMan by batches and ComBat by slides, either used alone or combined with median normalization, also offer fairly high C-indices ranging from 0.75 to 0.86. ComBat by batches, either used alone or with median normalization exhibits lower accuracy (C-index 0.72~0.83). Again, VSN and quantile normalization, either used alone or together with BatMan or ComBat, are the worst performers (C-index 0.69~0.76).

When batch effects are present and associated with survival time in the test data (BE11Cor01, BE11Cor11, and BE11Cor1-1), the performance of all methods deteriorates (C-index 0.57~0.77), with BatMan and median normalization being the best performers (C-index 0.72~0.77). Of note, when batch effects are negatively associated with survival time (BE1Cor1-1), all three normalization methods outperform BatMan and ComBat; however, this performance comes with the price of spurious predictive signals as we will show in the null-signal simulation scenario. Comparing across the eight panels, prediction accuracy is highest when there are no batch effects. It decreases for all methods when batch effects are present and further so when batch effects confound survival time. This overall trend reveals the chilling reality that batch effects negatively impact prediction modeling despite the use of data adjustment strategies. It also supports that BatMan is consistently a top performer, followed by ComBat and median normalization, while quantile normalization - currently the most popular strategy - does not satisfactorily remedy batch effects in survival prediction.

*Penalized regression method*

**Figure 2** displays the simulation results of LASSO-penalized Cox regression. The general patterns are very similar to those of the oracle method, with BatMan alone still a top performer across the eight simulation scenarios. A new observation is that, when batch effects are associated with survival time in the training data (BE10Cor10, BE11Cor10, BE11Cor01, BE11Cor11, and BE11Cor1-1), prediction accuracy is highly variable for quantile normalization and VSN as evidenced by exceptionally wide ranges between $2.5^{th}$ and $97.5^{th}$ percentiles of C-index, when used alone or in combination. This makes these methods even less desirable.

*Univariate selection method*

**Figure 3** presents the results when the univariate selection method is used to build the prediction model. In general, prediction accuracy becomes lower and substantially more variable across most adjustment strategies when compared to the oracle and LASSO methods. Notably, when batch effects are only present in training data and associated with survival time (BE10Cor10 and BE11Cor10), ComBat by batches (either alone or combined with median normalization) slightly outperforms BatMan (C-index 0.73~0.79 *versus* 0.72~0.76). However, as the univariate selection method results in lower prediction accuracy than the Lasso method across the board, the latter should be the preferred strategy for managing batch effects in practice.

### 3.2 Simulation with Weak Signals

Simulation results under weak signals are included in Supplementary Material (**S-Figures 2-5**). The general observations are consistent with those under moderate signals, demonstrating the robustness of our findings to the strength of prognostic signals.

### 3.3 Simulation with Null Signal

We further examine the performance of adjustment strategies under the null model where no feature is associated with the survival outcome. Note that the oracle method is not available under the null model. The results are presented in Supplementary Material (**S-Figures 6-8**). Under the null model, as expected, no modeling approach offers any predictive value regardless of the choice of adjustment strategies in general, which is reflected by C-index being around 0.5 (BE00Cor00, BE10Cor00, BE10Cor10, BE11Cor00, BE11Cor10, and BE11Cor01). However, when batch effects exist in both training and test data with positive association with the survival outcome (BE11Cor11), normalization and ComBat lead to small but notable predictive values beyond a random guess (C-index 0.52~0.56); conversely, when the association between batch effects and the survival outcome is negative (BE11Cor1-1), normalization and ComBat exhibit slightly predictive values in the reverse direction reflected by C-index being below 0.5 (C-index 0.45~0.48).

These observations suggest that batch effects that confound survival outcome can induce spurious predictive values, either positive or negative depending on the direction of the confounding association.

**3.4 Application to the TCGA Ovarian Cancer MiRNA Microarray Study**

The 466 samples in the TCGA study were collected in 98 batches and, as shown in **S-Figure 9** of the Supplementary Material, they exhibited strong evidence of batch effects. **Figure 4** depicts the average test-data C-indiex from the repeated splits of the TCGA ovarian cancer data. The prediction accuracy of BatMan exceeds that of ComBat for all three modeling approaches (univariate selection, Lasso, and adaptive Lasso) regardless of the use of normalization. Moreover, its performance is worsened by the addition of VSN or quantile normalization. These results are consistent with our findings in the simulation study.

## 4. DISCUSSION

While stratified regression is proven useful for analyzing low-dimensional data, it has been under-utilized for high-dimensional data and has not been applied for the purpose of managing batch effects in transcriptomics data. In this article, we propose to use stratification along with variable selection for simultaneously mitigating batch effects and building survival predictors. We demonstrated with both simulations and real data analysis that BatMan is more effective than ComBat, especially when there are batch effects in the data. As pointed out by Nygaard et al (2016) and Li & Johnson (2021), ComBat-adjusted data may possess biased group differences and inflated correlations among samples, leading to adverse impacts on downstream data analysis. In contrast,

BatMan cancels out batch effects by taking advantage of the partial likelihood structure and hence avoids estimating and subtracting batch effects.

The performance of BatMan depends on the size of the strata (that is, the batches). Our simulation study examines two stratum sizes, array slides (each of size eight) and array batches (each of size 32 to 40). Slides offer smaller stratum size than batches, which leads to lower efficiency and greater variance as well as a higher likelihood of having zero-event strata, especially when censoring rate is high. Stratification by slides outperforms by batches in our simulations, but this may not always be the case. Overall, there is a bias-variance trade-off in the choice of strata. In practice, one could use cross-validation to choose the batch variable that offers the best prediction. The performance of BatMan also depends on the use of a preceding normalization step. Adding such a step worsens the prediction especially when quantile normalization or variance stabilizing normalization is used, which is consistent with our recent study that evaluated the operating characteristics of data normalization for survival prediction (Ni and Qin, 2021).

We have previously showed that balanced study design in data collection (via the use of design elements such as blocking and randomization when assigning arrays to samples) is more effective than post-hoc data adjustment in the context of sample classification (Qin, Huang and Begg, 2016). For survival prediction, nested case-control designs can in principle be used for balancing batches and the outcome, where each case that incurs the event of interest is matched with one or more controls from the same batch to form a stratum; however, the issues of duplicate and empty matching inevitably arise in nested case-control design, making it not quite practical.

We simulate array data with virtual rehybridization and outcome data with permutation to avoid unrealistic parametric assumptions and to mimic empirical data as much as possible. The re-sampling-based methods help preserve marginal distributions and complex correlation structures in biologic variations and batch effects as well as the association between biologic signals and survival outcome. However, these methods are limited in their ability to single out the impact of one factor while holding the other factors constant. For example, they cannot isolate whether the negative impact of batch effects is due to the altered marginal distributions of features or the induced inter-feature correlations.

This article illustrates the benefits of the proposed BatMan method and assesses its performance along with ComBat and data normalization in the context of survival risk prediction. When developing a survival risk prediction model using transcriptomics data that possess batch effects, we recommend the use of BatMan without normalization. Even better, whenever possible, we encourage the use of uniform handling in data collection to prevent batch effects and generate high-quality data so that the data-to-knowledge translation can be accurate and reproducible.

# FUNDING

This work was supported by the National Institutes of Health [HG012124, CA214845, and CA008748].

# ACKNOWLEDGEMENTS

The authors thank Joseph Christoff for editorial assistance.

# SUPPLEMENTARY MATERIAL

Supplementary material is available online.

# REFERENCES

ACS (2022). Key Statistics for Ovarian Cancer. American Cancer Society.

AKAIKE, H. (1973). Maximum Likelihood Identification of Gaussian Autoregressive Moving Average Models. *Biometrika* **60**, 255-265.

AKBANI, R., ZHANG, N., BROOM, B. M. AND WEINSTEIN, J. N. (2022). TCGA Batch Effects Viewer. MD Anderson Cancer Center.

ALTMAN, D. (1990). *Practical statistics for medical research*. CRC press.

BARLIN, J. N., YU, C., HILL, E. K., ZIVANOVIC, O., KOLEV, V., LEVINE, D. A., SONODA, Y., ABU-RUSTUM, N. R., HUH, J., BARAKAT, R. R., KATTAN, M. W. AND CHI, D. S. (2012). Nomogram for predicting 5-year disease-specific mortality after primary surgery for epithelial ovarian cancer. *Gynecol Oncol* **125**, 25-30.

BARTEL, D. P. (2004). MicroRNAs: genomics, biogenesis, mechanism, and function. *Cell* **116**, 281-297.

BRADBURN, M. J., CLARK, T. G., LOVE, S. B. AND ALTMAN, D. G. (2003). Survival analysis part II: multivariate data analysis--an introduction to concepts and methods. *Br J Cancer* **89**, 431-436.

BROWN, A. W., KAISER, K. A. AND ALLISON, D. B. (2018). Issues with data and analyses: Errors, underlying themes, and potential solutions. *Proc Natl Acad Sci U S A* **115**, 2563-2570.

CANCER GENOME ATLAS RESEARCH, N. (2011). Integrated genomic analyses of ovarian carcinoma. *Nature* **474**, 609-615.

COCHRAN, W. G. (1968). The effectiveness of adjustment by subclassification in removing bias in observational studies. *Biometrics* **24**, 295-313.

CONSORTIUM, M., SHI, L., REID, L. H., JONES, W. D., SHIPPY, R., WARRINGTON, J. A., BAKER, S. C., COLLINS, P. J., DE LONGUEVILLE, F., KAWASAKI, E. S., LEE, K. Y., LUO, Y., SUN, Y. A., WILLEY, J. C., SETTERQUIST, R. A., FISCHER, G. M., TONG, W., DRAGAN, Y. P., DIX, D. J., FRUEH, F. W., GOODSAID, F. M., HERMAN, D., JENSEN, R. V., JOHNSON, C. D., LOBENHOFER, E. K., PURI, R. K., SCHRF, U., THIERRY-MIEG, J., WANG, C., WILSON, M., WOLBER, P. K., ZHANG, L., AMUR, S., BAO, W., BARBACIORU, C. C., LUCAS, A. B., BERTHOLET, V., BOYSEN, C., BROMLEY, B., BROWN, D., BRUNNER, A., CANALES, R., CAO, X. M., CEBULA, T. A., CHEN, J. J., CHENG, J., CHU, T. M., CHUDIN, E., CORSON, J., CORTON, J. C., CRONER, L. J., DAVIES, C., DAVISON, T. S., DELENSTARR, G., DENG, X., DORRIS, D., EKLUND, A. C., FAN, X. H., FANG, H., FULMER-SMENTEK, S., FUSCOE, J. C., GALLAGHER, K., GE, W., GUO, L., GUO, X., HAGER, J., HAJE, P. K., HAN, J., HAN, T., HARBOTTLE, H. C., HARRIS, S. C., HATCHWELL, E., HAUSER, C. A., HESTER, S., HONG, H., HURBAN, P., JACKSON, S. A., JI, H., KNIGHT, C. R., KUO, W. P., LECLERC, J. E., LEVY, S., LI, Q. Z., LIU, C., LIU, Y., LOMBARDI, M. J., MA, Y., MAGNUSON, S. R., MAQSODI, B., MCDANIEL, T., MEI, N., MYKLEBOST, O., NING, B., NOVORADOVSKAYA, N., ORR, M. S., OSBORN, T. W., PAPALLO, A., PATTERSON, T. A., PERKINS, R. G., PETERS, E. H., PETERSON, R., PHILIPS, K. L., PINE, P. S., PUSZTAI, L., QIAN, F., REN, H., ROSEN, M., ROSENZWEIG, B. A., SAMAHA, R. R., SCHENA, M., SCHROTH, G. P., SHCHEGROVA, S., SMITH, D. D., STAEDTLER, F., SU, Z., SUN, H., SZALLASI, Z., TEZAK, Z., THIERRY-MIEG, D., THOMPSON, K. L., TIKHONOVA, I., TURPAZ, Y., VALLANAT, B., VAN, C.,

WALKER, S. J., WANG, S. J., WANG, Y., WOLFINGER, R., WONG, A., WU, J., XIAO, C., XIE, Q., XU, J., YANG, W., ZHANG, L., ZHONG, S., ZONG, Y. AND SLIKKER, W., JR. (2006). The MicroArray Quality Control (MAQC) project shows inter- and intraplatform reproducibility of gene expression measurements. *Nat Biotechnol* **24**, 1151-1161.

CONSORTIUM, S. M.-I. (2014). A comprehensive assessment of RNA-seq accuracy, reproducibility and information content by the Sequencing Quality Control Consortium. *Nat Biotechnol* **32**, 903-914.

FAN, J., SAMWORTH, R. AND WU, Y. (2009). Ultrahigh dimensional feature selection: beyond the linear model. *J Mach Learn Res* **10**, 2013-2038.

FAN, J. Q. AND LV, J. C. (2008). Sure independence screening for ultrahigh dimensional feature space. *Journal of the Royal Statistical Society Series B-Statistical Methodology* **70**, 849-883.

GAGNON-BARTSCH, J. A. AND SPEED, T. P. (2012). Using control genes to correct for unwanted variation in microarray data. *Biostatistics* **13**, 539-552.

HAN, L., SAYYID, Z. N. AND ALTMAN, R. B. (2021). Modeling drug response using network-based personalized treatment prediction (NetPTP) with applications to inflammatory bowel disease. *PLoS Comput Biol* **17**, e1008631.

HARRELL, F. E., JR., LEE, K. L. AND MARK, D. B. (1996). Multivariable prognostic models: issues in developing models, evaluating assumptions and adequacy, and measuring and reducing errors. *Stat Med* **15**, 361-387.

HASTIE, T., TIBSHIRANI, R., FRIEDMAN, J. AND SPRINGERLINK (2009). *The Elements of Statistical Learning : Data Mining, Inference, and Prediction, Second Edition*. New York, NY: Springer New York : Imprint: Springer.

HUTTER, C. AND ZENKLUSEN, J. C. (2018). The Cancer Genome Atlas: Creating Lasting Value beyond Its Data. *Cell* **173**, 283-285.

IRIZARRY, R. A., HOBBS, B., COLLIN, F., BEAZER-BARCLAY, Y. D., ANTONELLIS, K. J., SCHERF, U. AND SPEED, T. P. (2003). Exploration, normalization, and summaries of high density oligonucleotide array probe level data. *Biostatistics* **4**, 249-264.

JOHNSON, W. E., LI, C. AND RABINOVIC, A. (2007). Adjusting batch effects in microarray expression data using empirical Bayes methods. *Biostatistics* **8**, 118-127.

KALBFLEISCH, J. D. AND PRENTICE, R. L. (2002). *The statistical analysis of failure time data*. Hoboken, N.J.: J. Wiley.

KERR, M. K., MARTIN, M. AND CHURCHILL, G. A. (2000). Analysis of variance for gene expression microarray data. *J Comput Biol* **7**, 819-837.

KRATZ, A. AND CARNINCI, P. (2014). The devil in the details of RNA-seq. *Nat Biotechnol* **32**, 882-884.

LAZAR, C., MEGANCK, S., TAMINAU, J., STEENHOFF, D., COLETTA, A., MOLTER, C., WEISS-SOLIS, D. Y., DUQUE, R., BERSINI, H. AND NOWE, A. (2013). Batch effect removal methods for microarray gene expression data integration: a survey. *Brief Bioinform* **14**, 469-490.

LEE, A. H. (2005). Prediction of cancer outcome with microarrays. *Lancet* **365**, 1685; author reply 1686.

LEEK, J. T., SCHARPF, R. B., BRAVO, H. C., SIMCHA, D., LANGMEAD, B., JOHNSON, W. E., GEMAN, D., BAGGERLY, K. AND IRIZARRY, R. A. (2010). Tackling the widespread and critical impact of batch effects in high-throughput data. *Nat Rev Genet* **11**, 733-739.

LEEK, J. T. AND STOREY, J. D. (2007). Capturing heterogeneity in gene expression studies by surrogate variable analysis. *PLoS Genet* **3**, 1724-1735.

LI, J., LENFERINK, A. E., DENG, Y., COLLINS, C., CUI, Q., PURISIMA, E. O., O'CONNOR-MCCOURT, M. D. AND WANG, E. (2010). Identification of high-quality cancer prognostic markers and metastasis network modules. *Nat Commun* **1**, 34.

LI, T., ZHANG, Y., PATIL, P. AND JOHNSON, W. E. (2021). Overcoming the impacts of two-step batch effect correction on gene expression estimation and inference. *Biostatistics*.

LOVEN, J., ORLANDO, D. A., SIGOVA, A. A., LIN, C. Y., RAHL, P. B., BURGE, C. B., LEVENS, D. L., LEE, T. I. AND YOUNG, R. A. (2012). Revisiting global gene expression analysis. *Cell* **151**, 476-482.

MCCALL, M. N., BOLSTAD, B. M. AND IRIZARRY, R. A. (2010). Frozen robust multiarray analysis (fRMA). *Biostatistics* **11**, 242-253.

MCKINNEY, E. F., CUTHBERTSON, I., HARRIS, K. M., SMILEK, D. E., CONNOR, C., MANFERRARI, G., CARR, E. J., ZAMVIL, S. S. AND SMITH, K. G. C. (2021). A CD8(+) NK cell transcriptomic signature associated with clinical outcome in relapsing remitting multiple sclerosis. *Nat Commun* **12**, 635.

MESTDAGH, P., HARTMANN, N., BAERISWYL, L., ANDREASEN, D., BERNARD, N., CHEN, C., CHEO, D., D'ANDRADE, P., DEMAYO, M., DENNIS, L., DERVEAUX, S., FENG, Y., FULMER-SMENTEK, S., GERSTMAYER, B., GOUFFON, J., GRIMLEY, C., LADER, E., LEE, K. Y., LUO, S., MOURITZEN, P., NARAYANAN, A., PATEL, S., PEIFFER, S., RUBERG, S., SCHROTH, G., SCHUSTER, D., SHAFFER, J. M., SHELTON, E. J., SILVERIA, S., ULMANELLA, U., VEERAMACHANENI, V., STAEDTLER, F., PETERS, T., GUETTOUCHE, T., WONG, L. AND VANDESOMPELE, J. (2014). Evaluation of quantitative miRNA expression platforms in the microRNA quality control (miRQC) study. *Nat Methods* **11**, 809-815.

NEKRUTENKO, A. AND TAYLOR, J. (2012). Next-generation sequencing data interpretation: enhancing reproducibility and accessibility. *Nat Rev Genet* **13**, 667-672.

NI, A. AND QIN, L. X. (2021). Performance evaluation of transcriptomics data normalization for survival risk prediction. *Brief Bioinform* **22**.

NYGAARD, V., RODLAND, E. A. AND HOVIG, E. (2016). Methods that remove batch effects while retaining group differences may lead to exaggerated confidence in downstream analyses. *Biostatistics* **17**, 29-39.

QIN, L. X., HUANG, H. C. AND BEGG, C. B. (2016). Cautionary Note on Using Cross-Validation for Molecular Classification. *J Clin Oncol* **34**, 3931-3938.

QIN, L. X., HUANG, H. C., VILLAFANIA, L., CAVATORE, M., OLVERA, N. AND LEVINE, D. A. (2018). A pair of datasets for microRNA expression profiling to examine the use of careful study design for assigning arrays to samples. *Sci Data* **5**, 180084.

QIN, L. X., HUANG, H. C. AND ZHOU, Q. (2014). Preprocessing Steps for Agilent MicroRNA Arrays: Does the Order Matter? *Cancer Inform* **13**, 105-109.

QIN, L. X. AND LEVINE, D. A. (2016). Study design and data analysis considerations for the discovery of prognostic molecular biomarkers: a case study of progression free survival in advanced serous ovarian cancer. *BMC Med Genomics* **9**, 27.

QIN, L. X. AND SATAGOPAN, J. M. (2009). Normalization method for transcriptional studies of heterogeneous samples--simultaneous array normalization and identification of equivalent expression. *Stat Appl Genet Mol Biol* **8**, Article 10.

QIN, L. X., ZHOU, Q., BOGOMOLNIY, F., VILLAFANIA, L., OLVERA, N., CAVATORE, M., SATAGOPAN, J. M., BEGG, C. B. AND LEVINE, D. A. (2014). Blocking and randomization to improve molecular biomarker discovery. *Clin Cancer Res* **20**, 3371-3378.

SIMON, N., FRIEDMAN, J., HASTIE, T. AND TIBSHIRANI, R. (2011). Regularization Paths for Cox's Proportional Hazards Model via Coordinate Descent. *J Stat Softw* **39**, 1-13.

TAMBA, C. L., NI, Y. L. AND ZHANG, Y. M. (2017). Iterative sure independence screening EM-Bayesian LASSO algorithm for multi-locus genome-wide association studies. *PLoS Comput Biol* **13**, e1005357.

TIBSHIRANI, R. (1997). The lasso method for variable selection in the Cox model. *Stat Med* **16**, 385-395.

VAN'T VEER, L. J. AND BERNARDS, R. (2008). Enabling personalized cancer medicine through analysis of gene-expression patterns. *Nature* **452**, 564-570.

WANG, C., ARMASU, S. M., KALLI, K. R., MAURER, M. J., HEINZEN, E. P., KEENEY, G. L., CLIBY, W. A., OBERG, A. L., KAUFMANN, S. H. AND GOODE, E. L. (2017). Pooled Clustering of High-Grade Serous Ovarian Cancer Gene Expression Leads to Novel Consensus Subtypes Associated with Survival and Surgical Outcomes. *Clin Cancer Res* **23**, 4077-4085.

ZHANG, Y., BERNAU, C., PARMIGIANI, G. AND WALDRON, L. (2020). The impact of different sources of heterogeneity on loss of accuracy from genomic prediction models. *Biostatistics* **21**, 253-268.

ZOU, H. (2006). The adaptive lasso and its oracle properties. *Journal of the American Statistical Association* **101**, 1418-1429.

**Table 1. Summary of simulation scenarios.** For scenario notation, 'BE' stands for Batch Effects; the first and second digits following 'BE' indicate presence ('1') *versus* absence ('0') of batch effects in the training and the test data, respectively; 'Cor' stands for correlation with survival outcome; The first and second digits following 'Cor' indicate the presence ('1'=positive, '-1'=negative) *versus* absence ('0') of correlation between batch effects and survival outcome in the training and test data, respectively.

| Scenario Notation | Batch effects in the training data | Batch effects in the test data | Batch effects correlated with outcome in the training data | Batch effects correlated with outcome in the test data |
|---|---|---|---|---|
| BE00Cor00 | No | No | No | No |
| BE10Cor00 | Yes | No | No | No |
| BE10Cor10 | Yes | No | Yes | No |
| BE11Cor00 | Yes | Yes | No | No |
| BE11Cor10 | Yes | Yes | Yes | No |
| BE11Cor01 | Yes | Yes | No | Yes |
| BE11Cor11 | Yes | Yes | Yes | Yes (positively) |
| BE11Cor1-1 | Yes | Yes | Yes | Yes (negatively) |

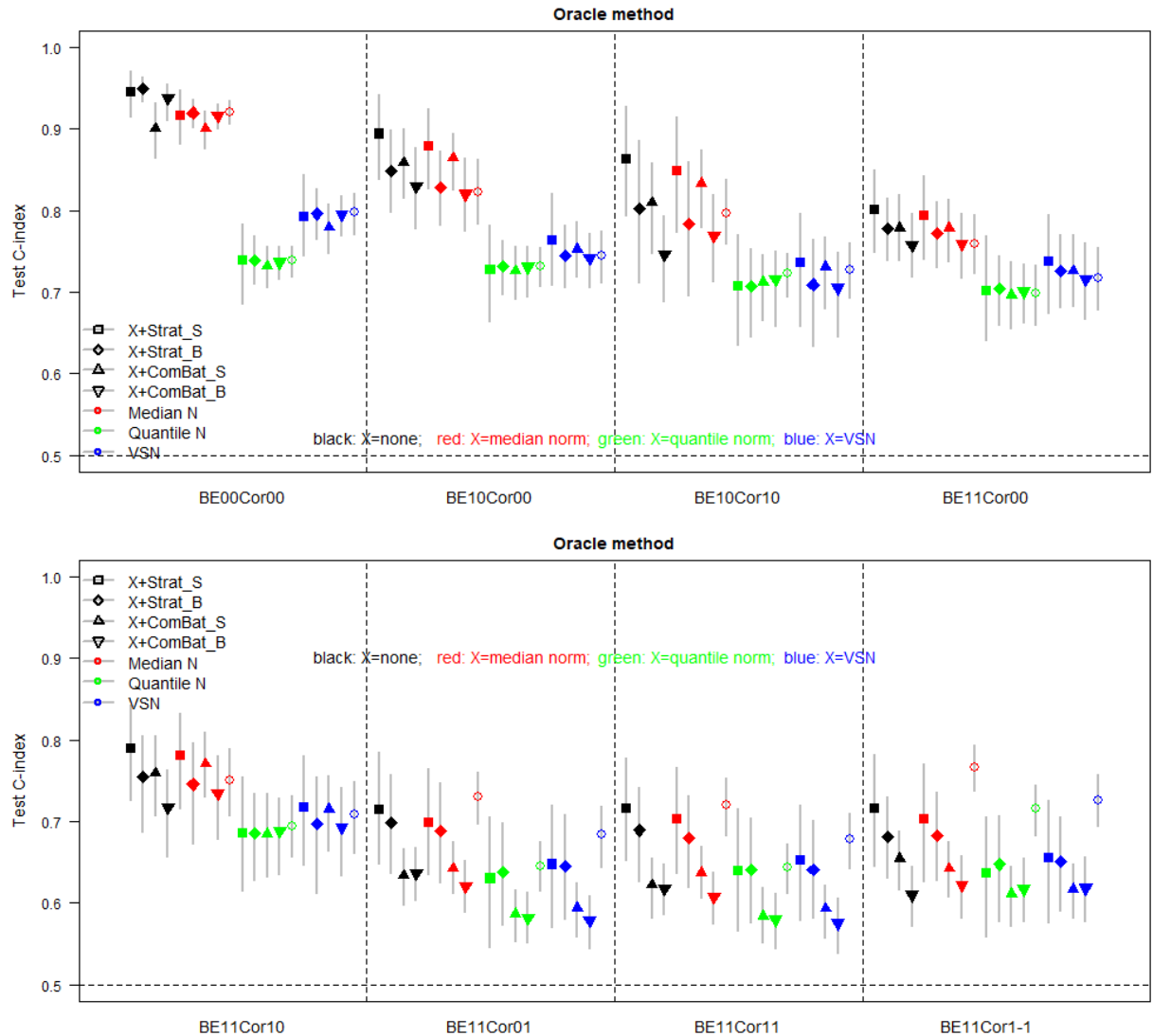

**Figure 1.** Test data Harrell's C-index of the prediction model developed by the oracle method under moderate signal. Vertical bars represent 2.5[th] and 97.5[th] percentiles. Symbols in the bars represent mean values. Symbol legend: *Strat_S* = stratification by array slides; *Strat_B* = stratification by slide batches; *ComBat_S* = ComBat by array slides; *ComBat_B* = ComBat by slide batches; *Median N* = median normalization; *Quantile N* = quantile normalization; *VSN* = variance stabilizing normalization. Panel legend: see notations in **Table 1**.

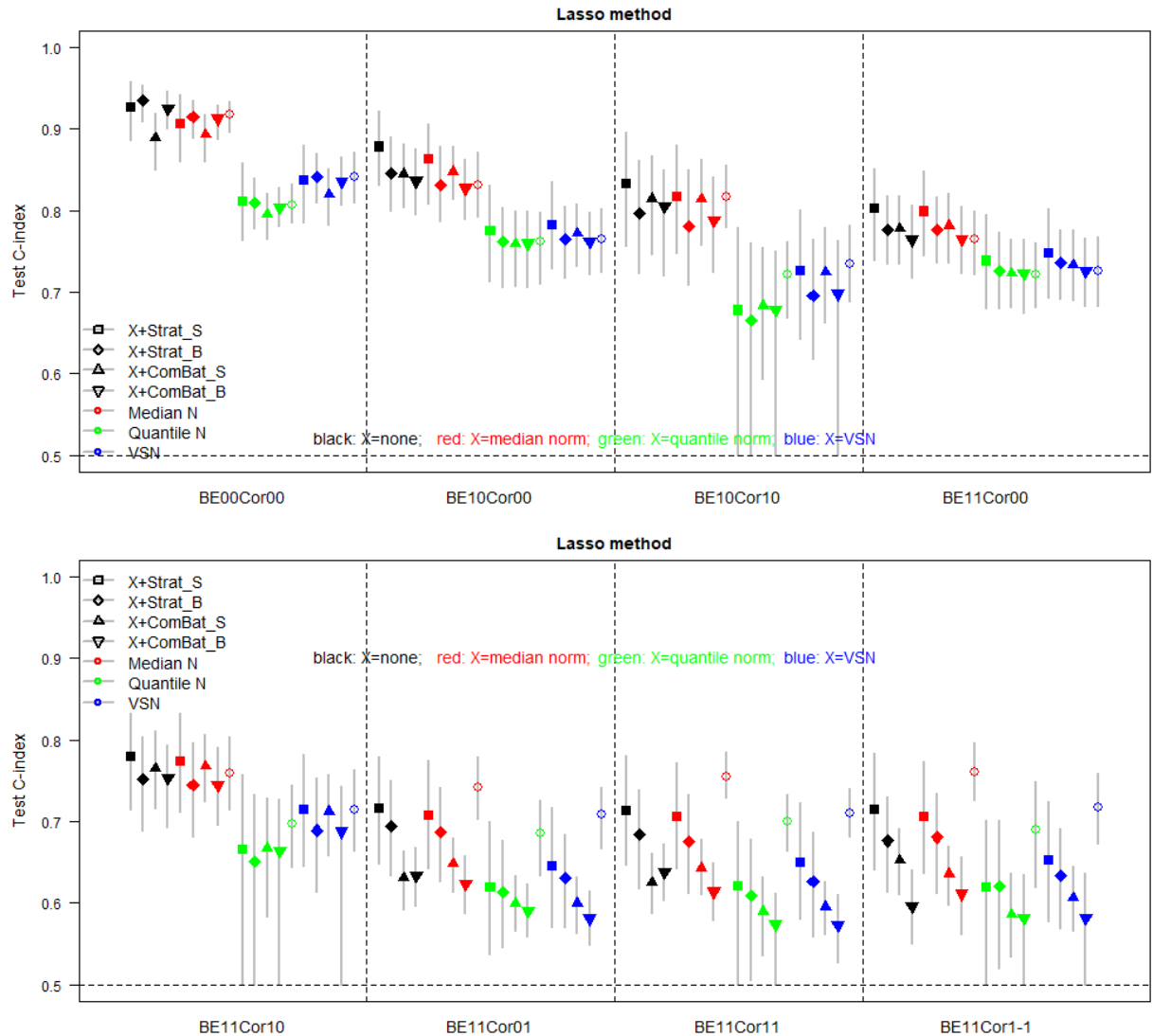

**Figure 2.** Test data Harrell's C-index of the prediction model developed by the Lasso method under moderate signal. Vertical bars represent 2.5$^{th}$ and 97.5$^{th}$ percentiles. Symbols in the bars represent mean values. Symbol legend: *Strat_S* = stratification by array slides; *Strat_B* = stratification by slide batches; *ComBat_S* = ComBat by array slides; *ComBat_B* = ComBat by slide batches; *Median N* = median normalization; *Quantile N* = quantile normalization; *VSN* = variance stabilizing normalization. Panel legend: see notations in **Table 1**.

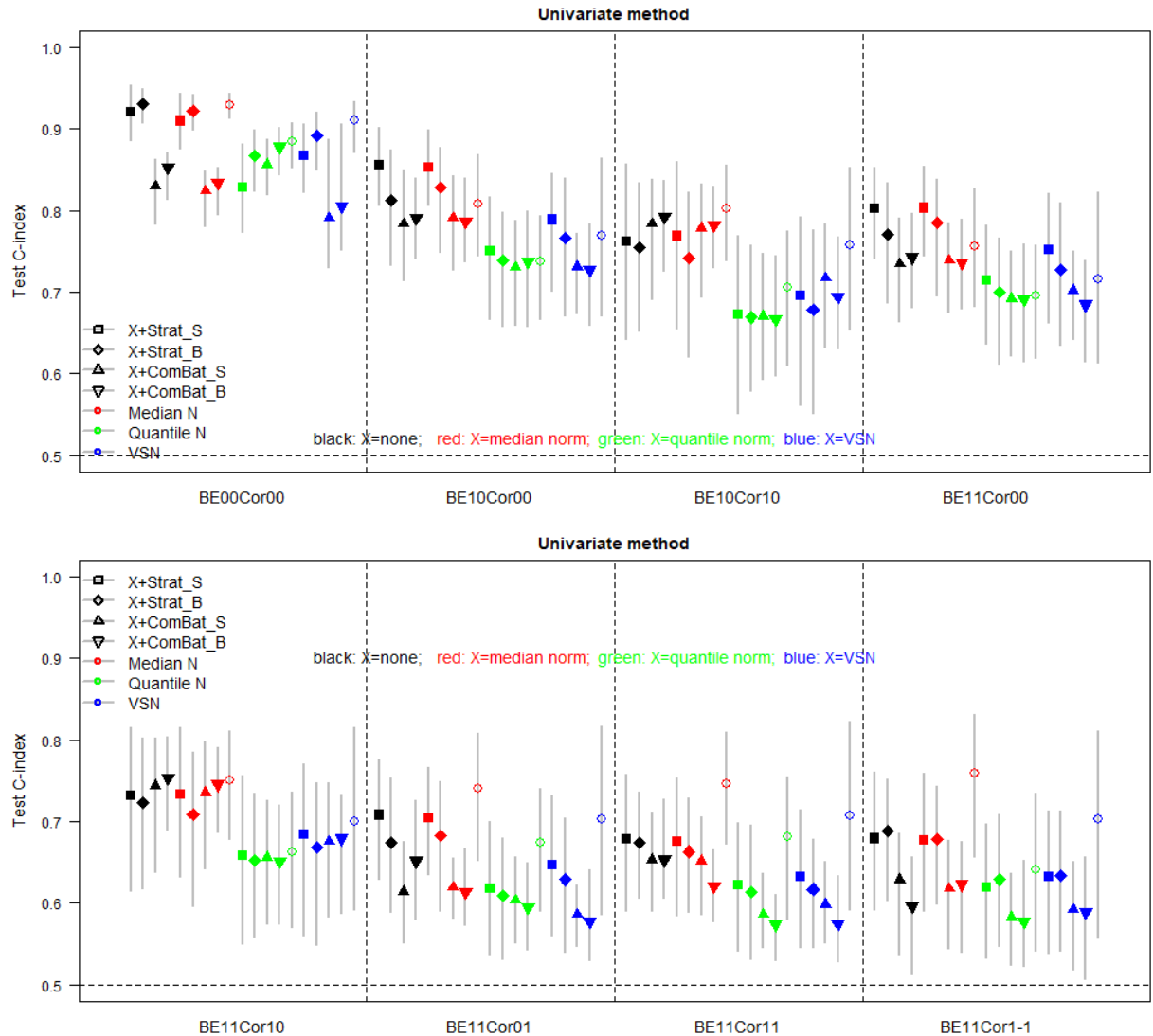

**Figure 3.** Test data Harrell's C-index of the prediction model developed by the univariate method under moderate signal. Vertical bars represent 2.5[th] and 97.5[th] percentiles. Symbols in the bars represent mean values. Symbol legend: *Strat_S* = stratification by array slides; *Strat_B* = stratification by slide batches; *ComBat_S* = ComBat by array slides; *ComBat_B* = ComBat by slide batches; *Median N* = median normalization; *Quantile N* = quantile normalization; *VSN* = variance stabilizing normalization. Panel legend: see notations in **Table 1**.

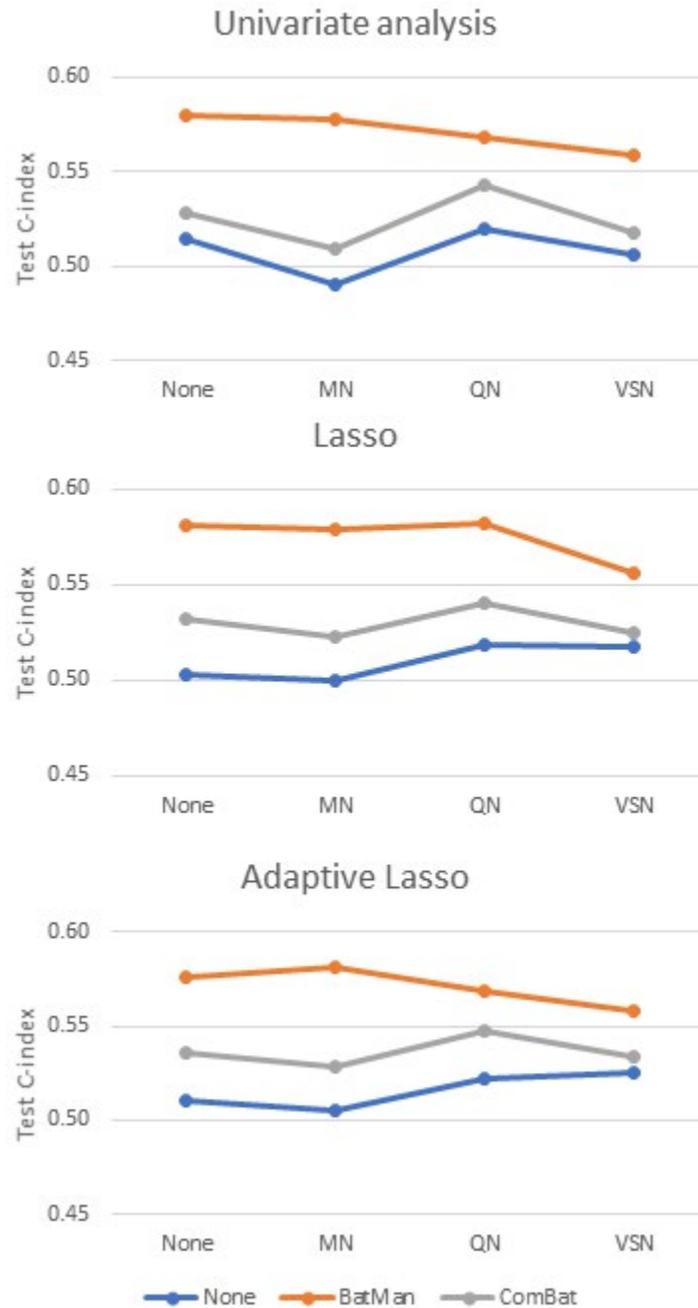

**Figure 4.** Test data Harrell's C-index of the prediction model developed based on the TCGA ovarian cancer microRNA microarray data. Dots represent the average test data C-index from the 50 repeated random splitting of the original dataset. *MN* = median normalization; *QN* = quantile normalization; *VSN* = variance stabilizing normalization**.**

**SUPPLEMENTS**

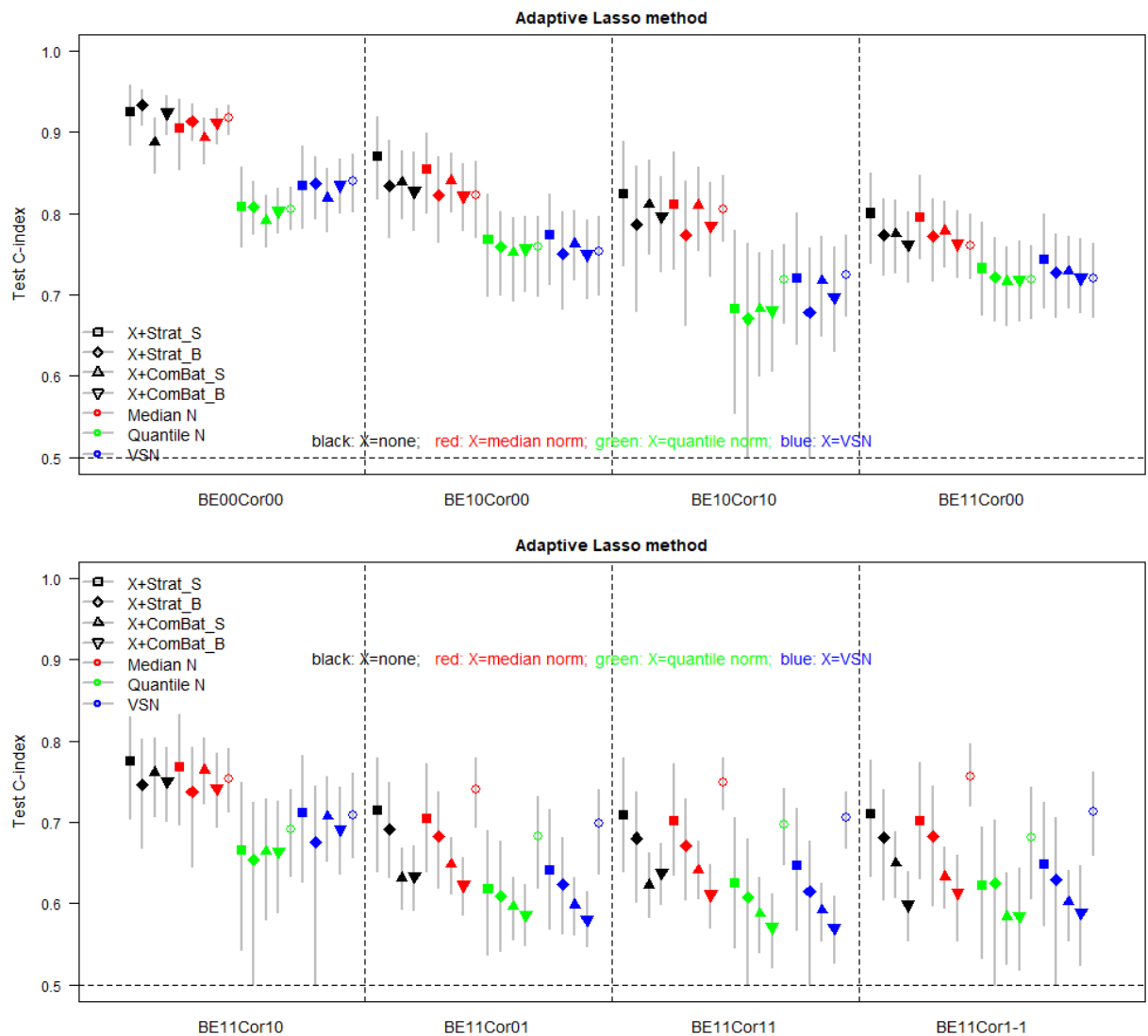

**S-Figure 1.** Test data Harrell's C-index of the prediction model developed by the adaptive Lasso method under moderate signal. Vertical bars represent 2.5[th] and 97.5[th] percentiles. Symbols in the bars represent mean values. Symbol legend: *Strat_S* = stratification by array slides; *Strat_B* = stratification by slide batches; *ComBat_S* = ComBat by array slides; *ComBat_B* = ComBat by slide batches; *Median N* = median normalization; *Quantile N* = quantile normalization; *VSN* = variance stabilizing normalization. Panel legend: see notations in **Table 1**.

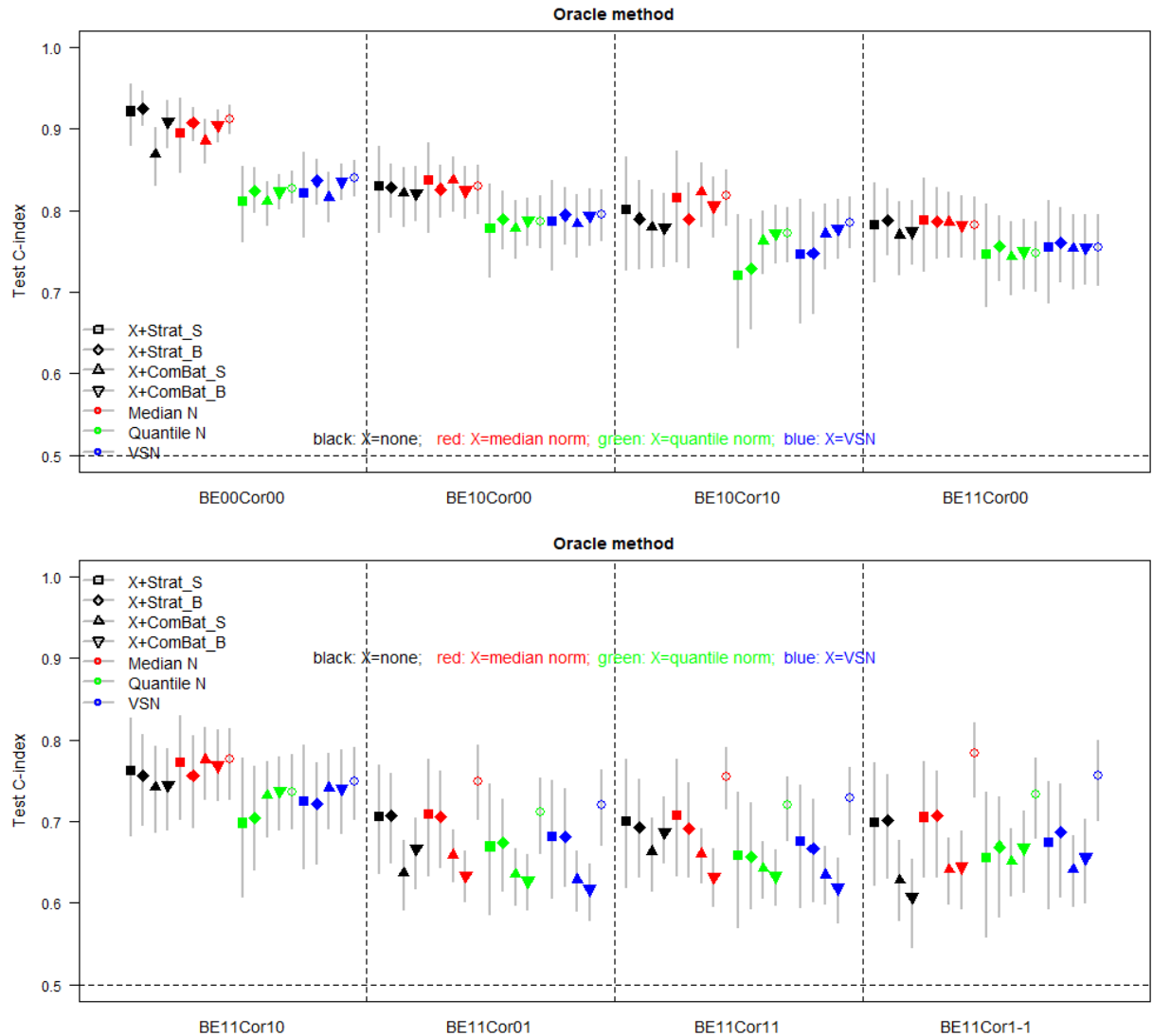

**S-Figure 2.** Test data Harrell's C-index of the prediction model developed by the oracle method under small signal. Vertical bars represent $2.5^{th}$ and $97.5^{th}$ percentiles. Symbols in the bars represent mean values. Symbol legend: *Strat_S* = stratification by array slides; *Strat_B* = stratification by slide batches; *ComBat_S* = ComBat by array slides; *ComBat_B* = ComBat by slide batches; *Median N* = median normalization; *Quantile N* = quantile normalization; *VSN* = variance stabilizing normalization. Panel legend: see notations in **Table 1**.

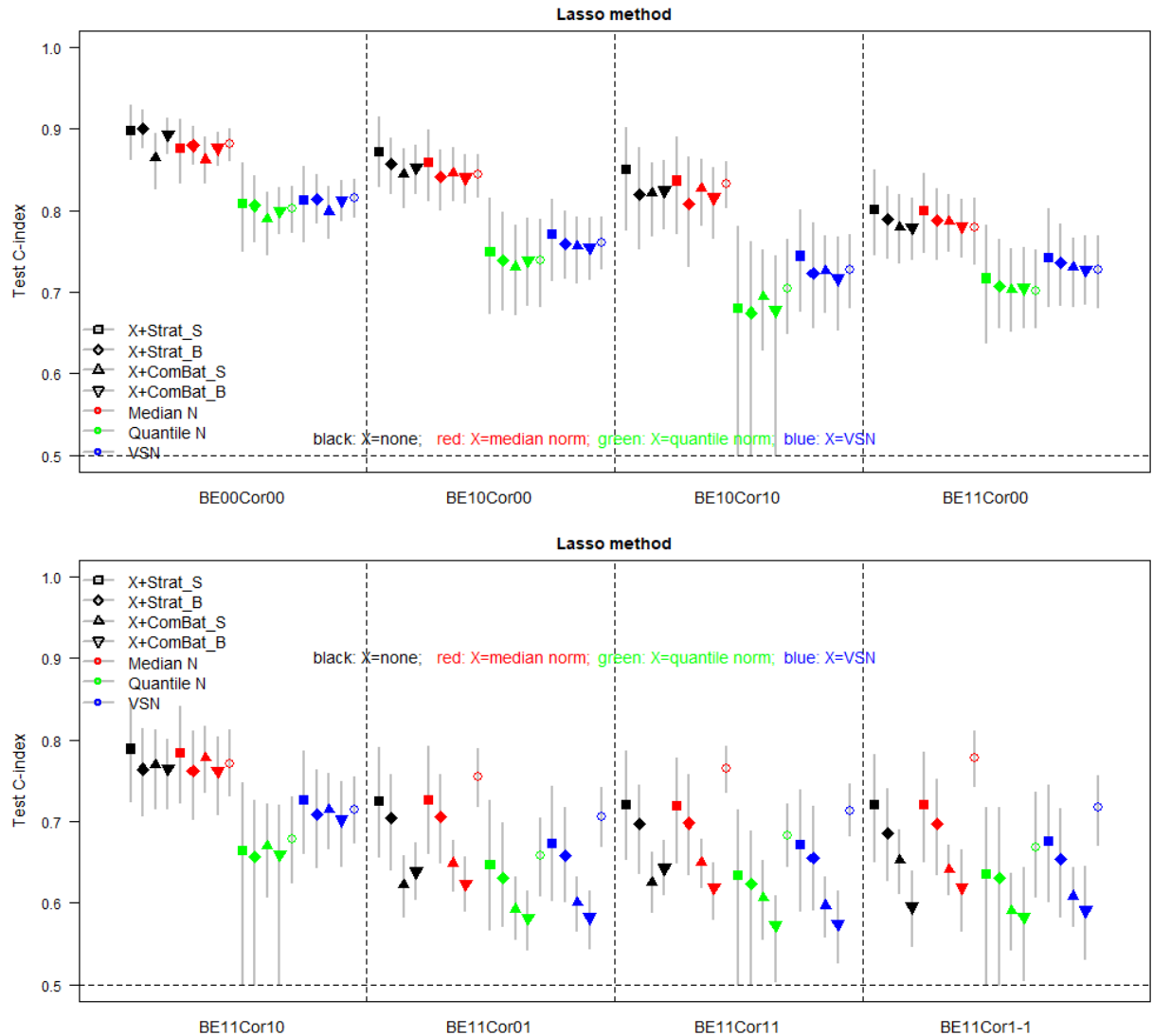

**S-Figure 3.** Test data Harrell's C-index of the prediction model developed by the Lasso method under small signal. Vertical bars represent $2.5^{th}$ and $97.5^{th}$ percentiles. Symbols in the bars represent mean values. Symbol legend: *Strat_S* = stratification by array slides; *Strat_B* = stratification by slide batches; *ComBat_S* = ComBat by array slides; *ComBat_B* = ComBat by slide batches; *Median N* = median normalization; *Quantile N* = quantile normalization; *VSN* = variance stabilizing normalization. Panel legend: see notations in **Table 1**.

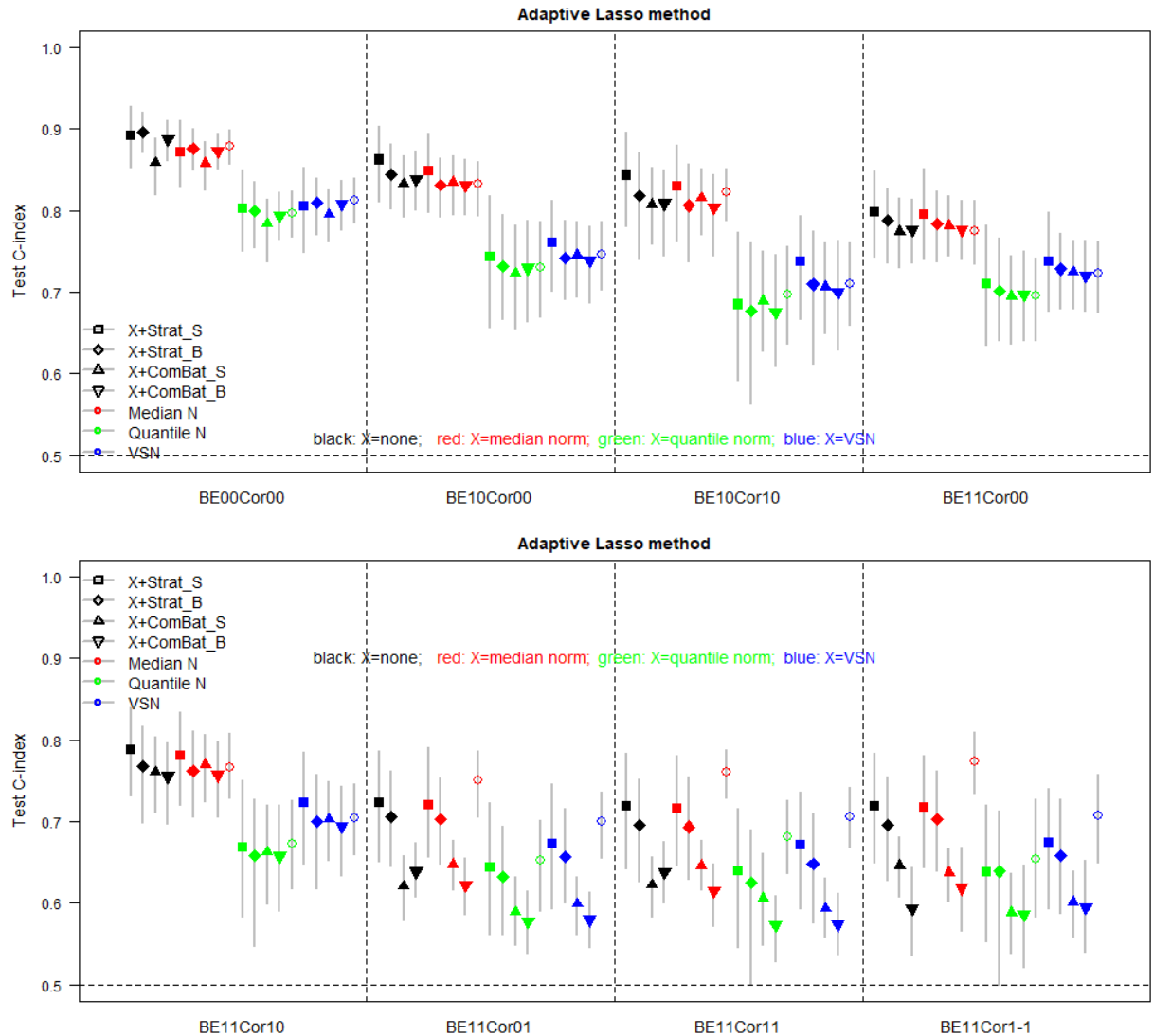

**S-Figure 4.** Test data Harrell's C-index of the prediction model developed by the adaptive Lasso method under small signal. Vertical bars represent 2.5$^{th}$ and 97.5$^{th}$ percentiles. Symbols in the bars represent mean values. Symbol legend: *Strat_S* = stratification by array slides; *Strat_B* = stratification by slide batches; *ComBat_S* = ComBat by array slides; *ComBat_B* = ComBat by slide batches; *Median N* = median normalization; *Quantile N* = quantile normalization; *VSN* = variance stabilizing normalization. Panel legend: see notations in **Table 1**.

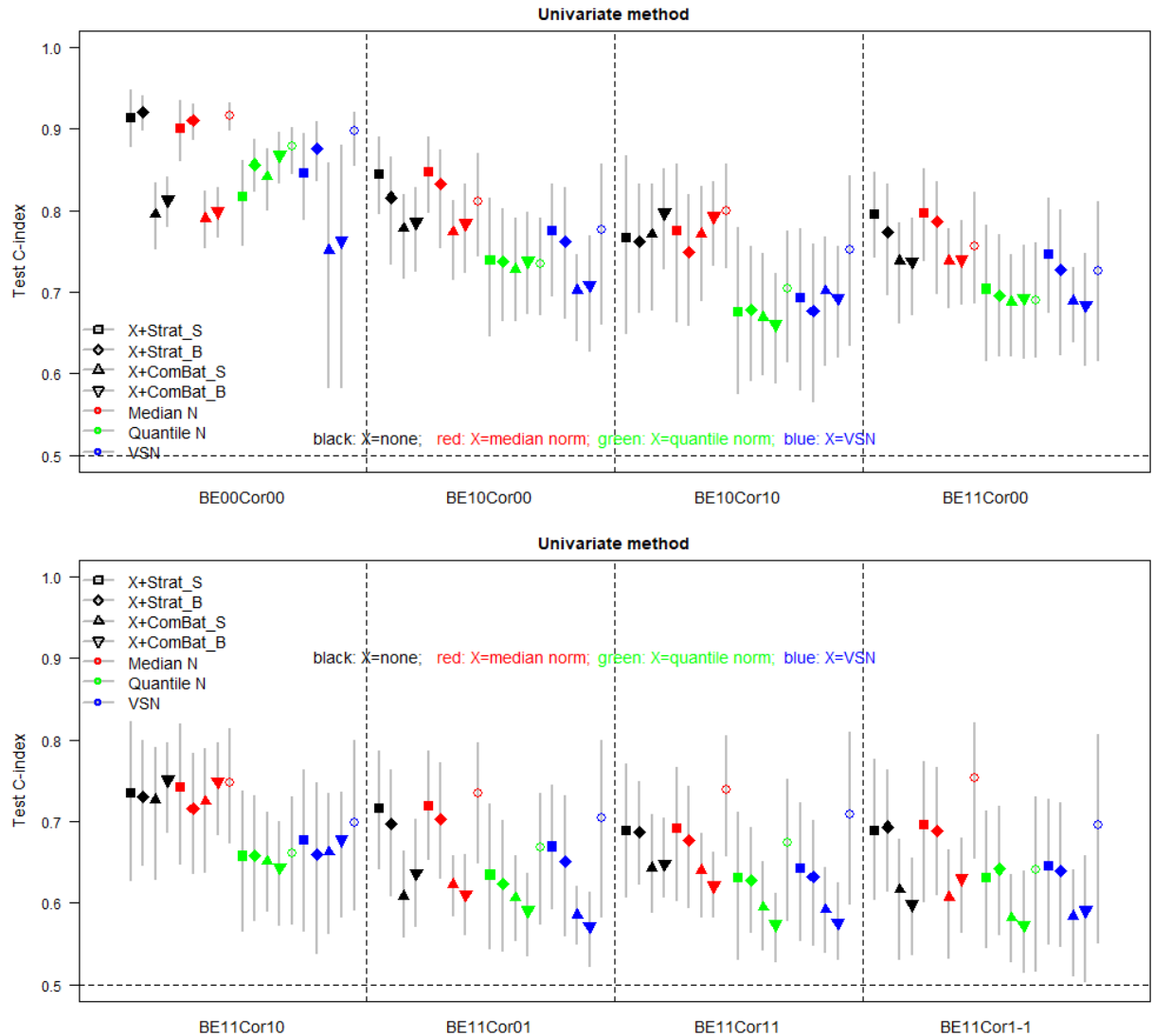

**S-Figure 5.** Test data Harrell's C-index of the prediction model developed by the univariate method under small signal. Vertical bars represent 2.5[th] and 97.5[th] percentiles. Symbols in the bars represent mean values. Symbol legend: *Strat_S* = stratification by array slides; *Strat_B* = stratification by slide batches; *ComBat_S* = ComBat by array slides; *ComBat_B* = ComBat by slide batches; *Median N* = median normalization; *Quantile N* = quantile normalization; *VSN* = variance stabilizing normalization. Panel legend: see notations in **Table 1**.

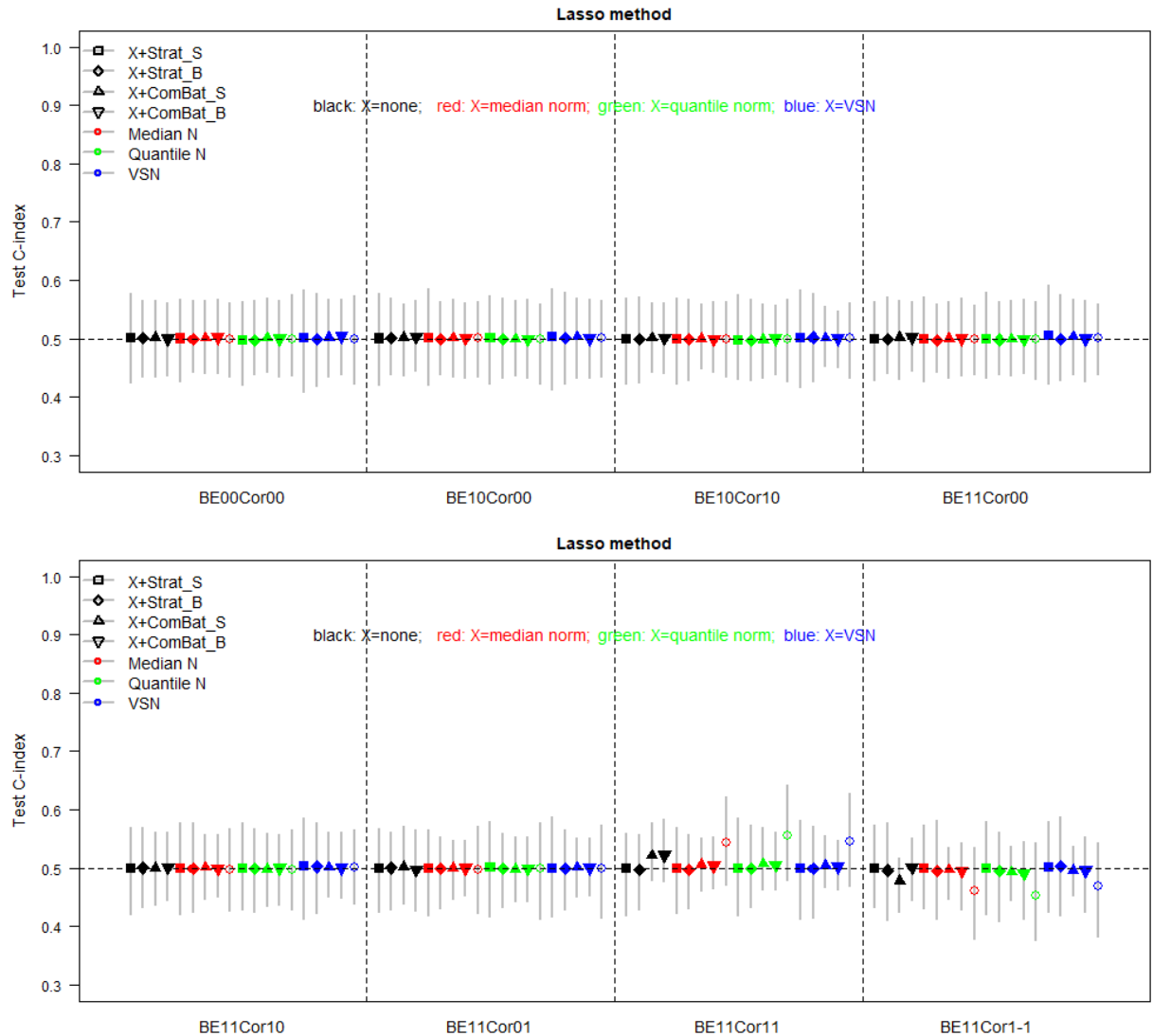

**S-Figure 6.** Test data Harrell's C-index of the prediction model developed by the Lasso method under null signal. Vertical bars represent 2.5[th] and 97.5[th] percentiles. Symbols in the bars represent mean values. Symbol legend: *Strat_S* = stratification by array slides; *Strat_B* = stratification by slide batches; *ComBat_S* = ComBat by array slides; *ComBat_B* = ComBat by slide batches; *Median N* = median normalization; *Quantile N* = quantile normalization; *VSN* = variance stabilizing normalization. Panel legend: see notations in **Table 1**.

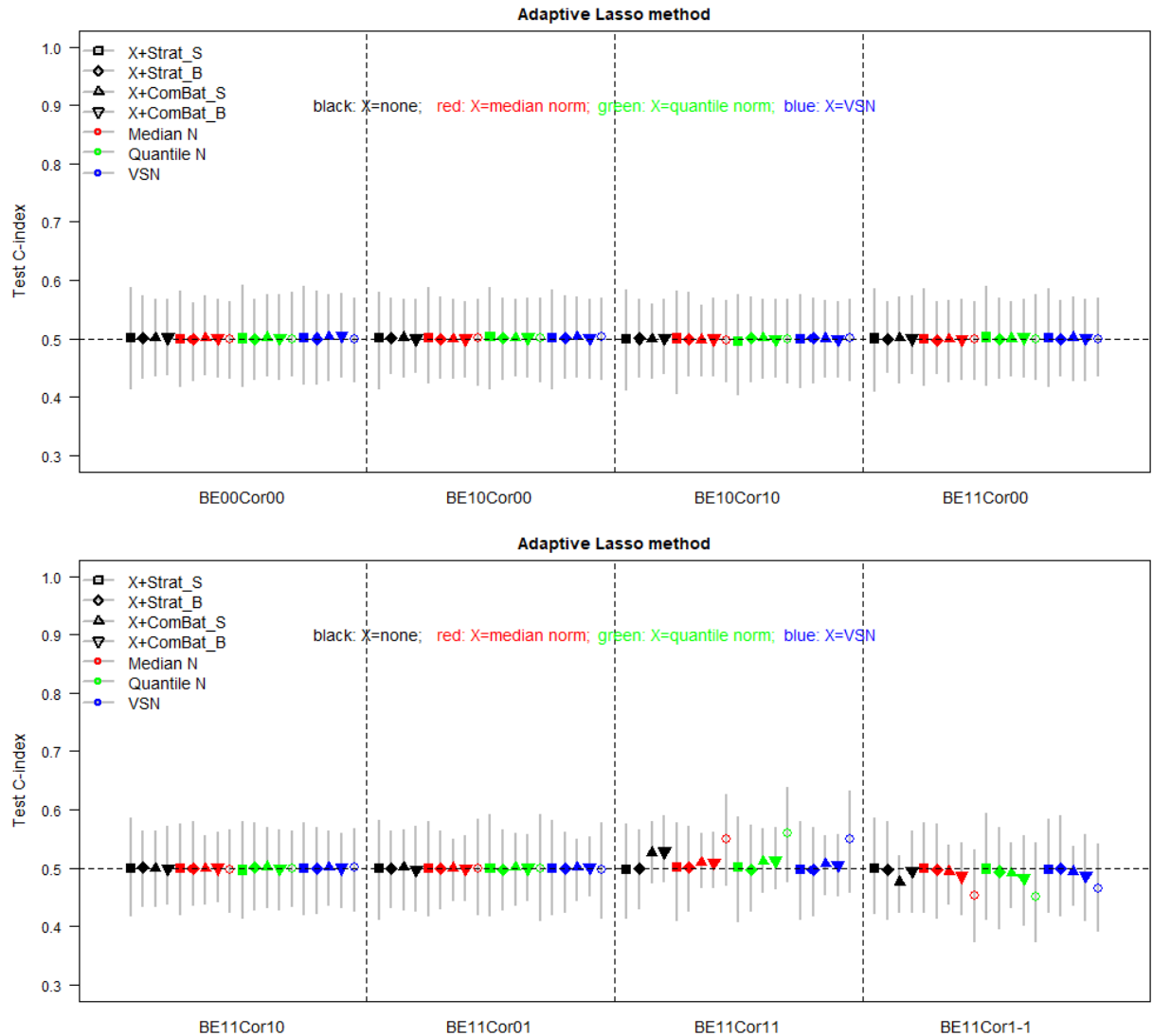

**S-Figure 7.** Test data Harrell's C-index of the prediction model developed by the adaptive Lasso method under null signal. Vertical bars represent 2.5[th] and 97.5[th] percentiles. Symbols in the bars represent mean values. Symbol legend: *Strat_S* = stratification by array slides; *Strat_B* = stratification by slide batches; *ComBat_S* = ComBat by array slides; *ComBat_B* = ComBat by slide batches; *Median N* = median normalization; *Quantile N* = quantile normalization; *VSN* = variance stabilizing normalization. Panel legend: see notations in **Table 1**.

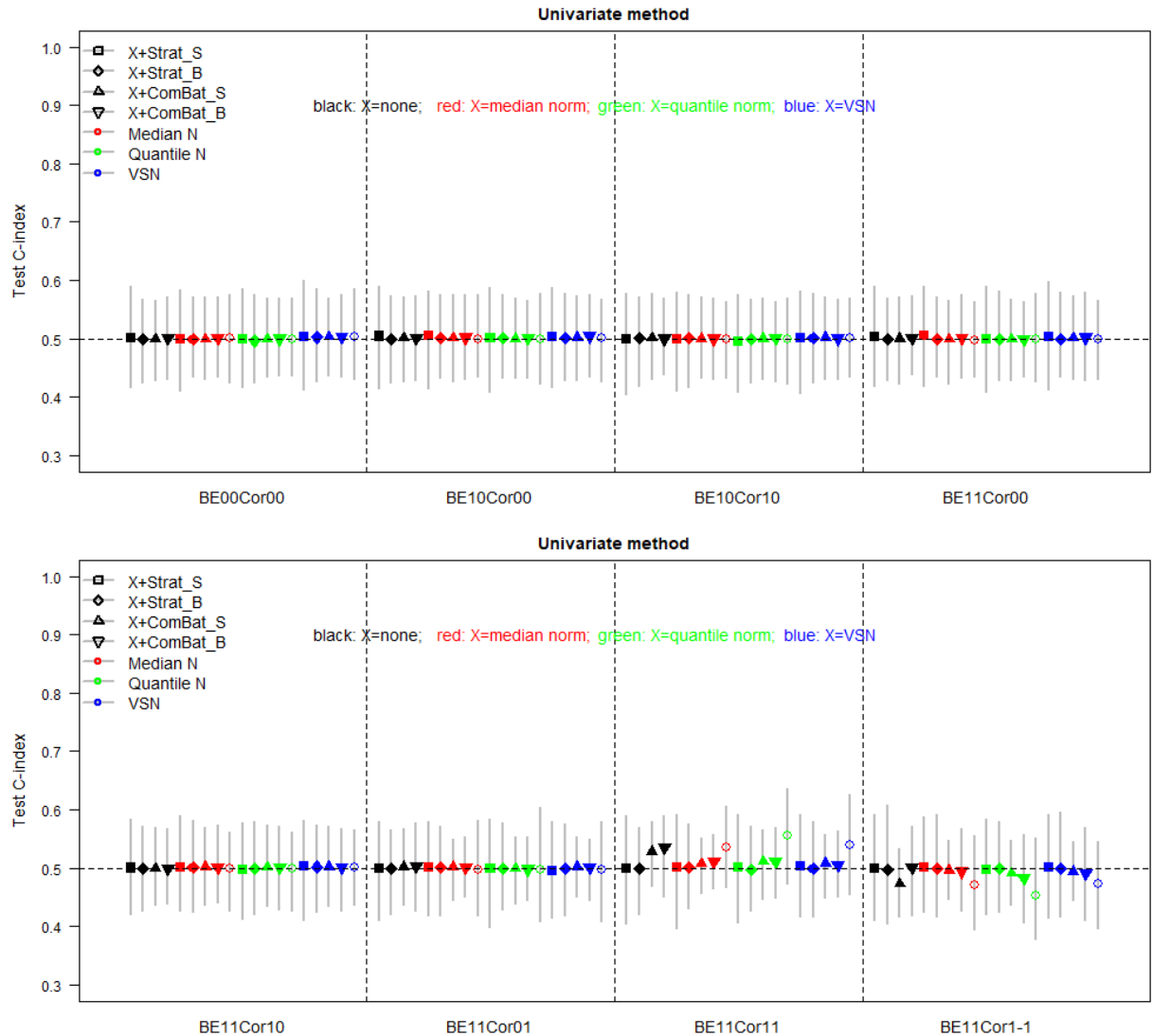

**S-Figure 8.** Test data Harrell's C-index of the prediction model developed by the univariate method under null signal. Vertical bars represent 2.5[th] and 97.5[th] percentiles. Symbols in the bars represent mean values. Symbol legend: *Strat_S* = stratification by array slides; *Strat_B* = stratification by slide batches; *ComBat_S* = ComBat by array slides; *ComBat_B* = ComBat by slide batches; *Median N* = median normalization; *Quantile N* = quantile normalization; *VSN* = variance stabilizing normalization. Panel legend: see notations in **Table 1**.

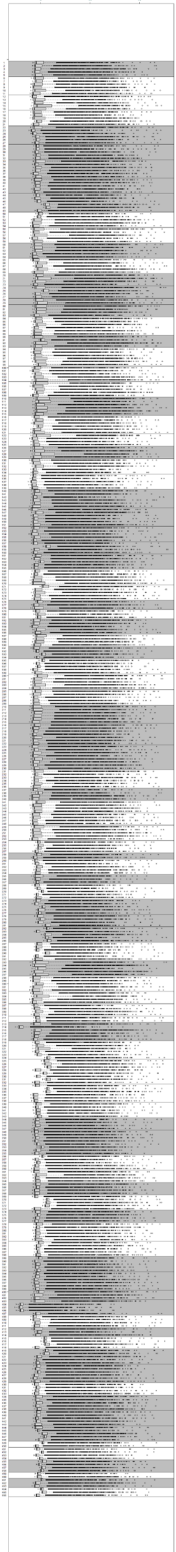

**S-Figure 9.** Boxplots of miRNA microarray data for the 466 samples in the TCGA ovarian cancer study. Each boxplot represents a tumor sample. Alternating shading and non-shading indicate experimental batches.